\documentclass{iau-JDSS}
\usepackage{graphicx,natbib}

\newcommand{\Lbol}{\mbox{$L_{\rm bol}$}}

\title[SpS 7.~~Testing Subtellar Models] 
{Testing Models with Brown Dwarf Binaries}

\author[Trent J. Dupuy \& Michael C. Liu]   
{Trent J. Dupuy \& Michael C. Liu}

\affiliation{ Institute for Astronomy, University of Hawai`i \\
  2680 Woodlawn Drive, Honolulu, HI, USA \\
}

\pubyear{20XX}
\volume{xxx}  
\pagerange{1}
\jname{Young Stars, Brown Dwarfs, and Protoplanetary Disks}
\editors{J.\ Gregorio-Hetem \& S.\ Alencar eds.}
\begin{document}

\maketitle

\begin{abstract}

  We have been using Keck laser guide star adaptive optics to monitor
  the orbits of ultracool binaries, providing dynamical masses at
  lower luminosities and temperatures than previously available and
  enabling strong tests of theoretical models.  (1) We find that model
  color--magnitude diagrams cannot reliably be used to infer masses as
  they do not accurately reproduce the colors of ultracool dwarfs of
  known mass.  (2) Effective temperatures inferred from evolutionary
  model radii can be inconsistent with temperatures derived from
  fitting observed spectra with atmospheric models by at most
  100--300~K.  (3) For the single pair of field brown dwarfs with a
  precise mass (3\%) \emph{and} age determination ($\approx$25\%), the
  measured luminosities are $\sim$2--3$\times$ higher than predicted
  by model cooling rates (masses inferred from \Lbol\ and age are
  20--30\% larger than measured).  Finally, as the sample of binaries
  with measured orbits grows, novel tests of brown dwarf formation
  theories are made possible (e.g., testing theoretical eccentricity
  distributions).


\end{abstract}

\begin{figure}[h]
  \begin{center}
    \vskip -0.15 in
    \hskip -0.1 in
    \includegraphics[width=2.1in,angle=0]{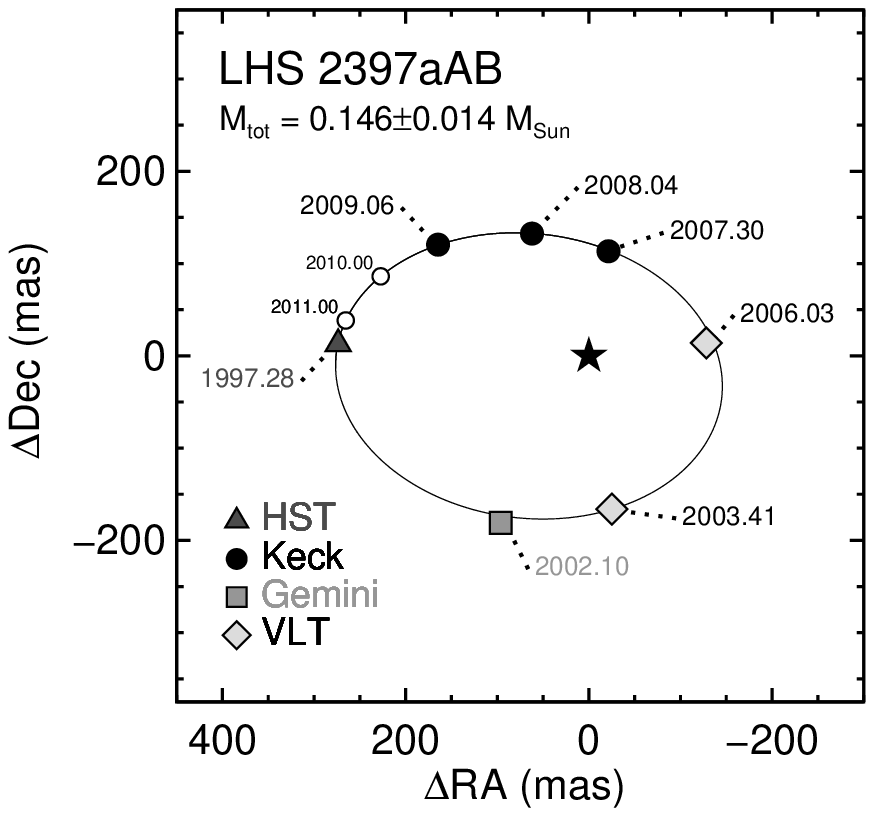} 
    \hskip  0.2 in
    \includegraphics[width=2.1in,angle=0]{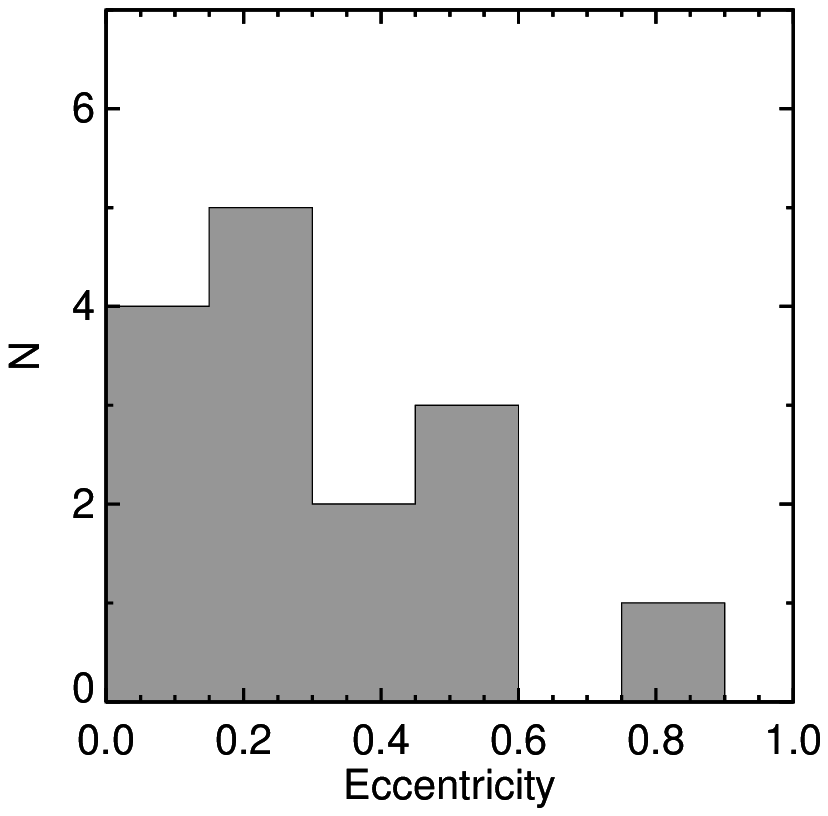} 
    \caption{\textit{Left:} Orbit of the M8+L7 binary LHS~2397aAB
      \citep{me-2397a}.  \textit{Right:} Eccentricity distribution for
      all nine published ultracool ($>$M6) binary orbits
      \citep{2004A&A...424..213B, 2006ApJ...644.1183S,
        2008A&A...484..429S, 2008ApJ...689..436L, 2008ApJ...678L.125B,
        2009AIPC.1094..509C, 2009ApJ...692..729D, me-2397a, me-2206},
      along with six unpublished orbits known to the author.}
    \label{fig1}
  \end{center}
\end{figure}

\newcommand\apj{\textit{ApJ}}
\newcommand\apjl{\textit{Ap. Lett.}}
\newcommand\aap{\textit{A\&A}}


\end{document}